# SIMULATION OF A TIME-OF-FLIGHT TELESCOPE FOR SUPRATHERMAL IONS IN THE HELIOSPHERE

R. BUČÍK, A. KORTH, U. MALL

*Max-Planck-Institut für Sonnensystemforschung*
*37191 Katlenburg-Lindau, Germany*

G. M. MASON

*JHU, Applied Physics Laboratory*
*Laurel, MD 20723, USA*

A Monte Carlo code based on Geant 3.21 has been used for simulations of energy losses and angular scattering in a time-of-flight Suprathermal Ion Telescope (SIT) on the Solar-Terrestrial Relations Observatory (STEREO). A hemispherical isotropic particle distribution, a monoenergetic or power law in energy is used in these simulations. The impact of scattering, energy losses and system noise on the instrument mass resolution is discussed.

## 1. Introduction

A Monte Carlo simulation of the SIT telescope aboard the STEREO spacecraft was performed by Geant 3.21 [1]. This is a powerful code which allows to simulate the whole experimental setup and tracks the particle within the setup taking into account the primary interaction of those particles with matter.

In the present paper we study effects of the angular scattering and energy losses in the time-of-flight (TOF) telescope on the mass resolution. Monte Carlo simulations are needed to understand the response of the SIT instrument, especially for masses and energies not covered by calibration measurements.

## 2. Experiment

The SIT sensor is a TOF mass spectrometer, which measures ions (from hydrogen up to iron) from ~20 keV/n to several MeV/n [2]. The instrument identifies the incident ion mass and the energy by measuring the TOF $\tau$ and the residual kinetic energy $E_{SSD}$ of the particle that enters the telescope through a thin nickel foil and stops in the silicon solid state detector (SSD) at the rear of the telescope. The TOF is determined by start and stop pulses from micro channel plates (MCPs) that detect secondary electrons that are emitted from the foil and the surface of the SSD. The mass $m$ of the ion is obtained from:

$$m = 2E_{SSD}\left(\frac{\tau}{L}\right)^2, \qquad (1)$$

where $L$ is the distance between the foil and the SSD. In the range ~ 0.1 -1.0 MeV/n, the mass resolution $\sigma/m$ is 0.1 measured during calibrations by α-sources. The TOF for SIT ranges from ~ 2.5 to 125 ns. The SSD threshold energy is set at 0.24 MeV and the total energy deposit is 163 MeV.

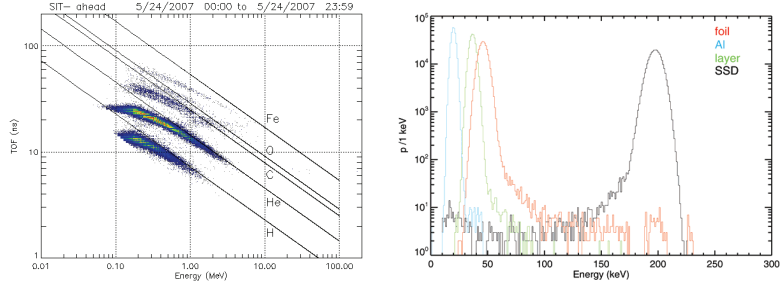

Figure 1. (Left) TOF versus the total kinetic energy for particles measured by the SIT sensor. Overplotted are straight lines for five species using Eq. (1). (Right) Energy loss histograms for 300 keV protons for different absorbers in the SIT telescope.

As can be seen in Eq. (1), when $E_{SSD}$ is plotted versus $\tau$ on a log-log scale, the various atomic species are organized along straight lines with slopes of (-2) and offsets given by the mass. This can be seen in Figure 1 (left), which shows data from the SIT sensor. Each point represents the measurements of one ion.

## 3. Overview of the simulations

The simple mass model of the SIT telescope consists of four absorbers: the two nickel entrance foils, each 1000 Å, are replaced by one 40.7x15.5 mm nickel foil, 0.2 μm thick; a 40x15 mm silicon SSD, with thickness of 500 μm, at a distance of $L$=10 cm behind the foil; adjacent to the SSD an aluminum surface metallization, 0.2 μm thick and a silicon junction dead layer, 0.35 μm thick. The MCPs rates are not implemented in the simulations, only the foil and SSD responses are taken into consideration. The telescope housing was not modeled; we neglected any additional effects scattering off the surrounding structure. The sunshade with its additional role of limiting the solid angle was not included in model geometry.

In our analysis $10^8$ particles were simulated which isotropically incident on SIT from the upper hemisphere. The simulated species include the following fully ionized ions: H, $^3$He, $^4$He, C, N, and O. The species are monoenergetic or with a power law spectrum with an index of (-2) for a total kinetic energy between 240 keV and 163 MeV. Except for ionization energy loss, simulations include energy straggling and multiple Coulomb scattering. The non-ionizing nuclear energy losses important for <100 keV/n heavy ions [3], are not considered in the present simulations.

## 4. Results and discussion

### 4.1. *Angular and energy distribution*

Energy losses in the four absorbers are given for 300 keV protons in the right panel of Figure 1. As shown later, the high energy losses in the foil correspond to high incident angles. The low energy tail in the SSD is a result of reduced ion energy by the absorbers above the SSD. On the average, 15% (3%) energy is deposited in the foil by 300 (900) keV protons. The same amount of energy is stored in the adjacent to the SSD Al metallization and dead layer.

Figure 2 (left) is an example of the number distribution of 300 keV protons as a function of the energy loss in the SSD and the angle between incident direction and normal of the foil. We see that particles with incident angles outside of the nominal telescope field-of-view (44° x 17°) can hit the SSD. In this figure lower energy deposit in the SSD corresponds to higher incident angles, which can be explained by a high energy lost in the foil, when a particle strikes it in a high incident angle and traverses through higher amount of material. Figure 2 (right) shows the distribution of a number of protons as a function of energy deposited in the SSD and the angle between incident and scattering direction. Particle which hit the SSD can be scattered to a high degree only for large incident angles. Those particles, however, store small amount of energy in the SSD. This fits well with the simulations.

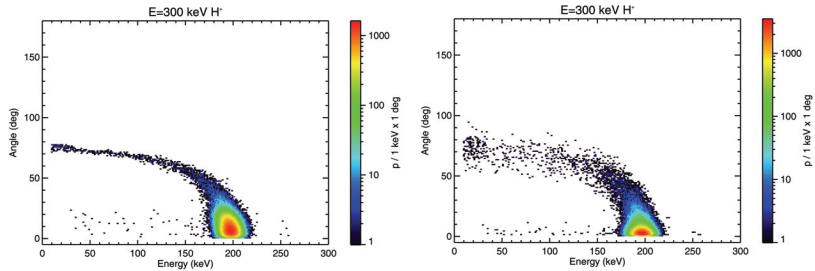

Figure 2. The simulated number distribution of 300 keV protons as a function of energy stored in the SSD and angle between incident direction and normal of the entrance foil (Left) or angle between incident and scattering direction (Right).

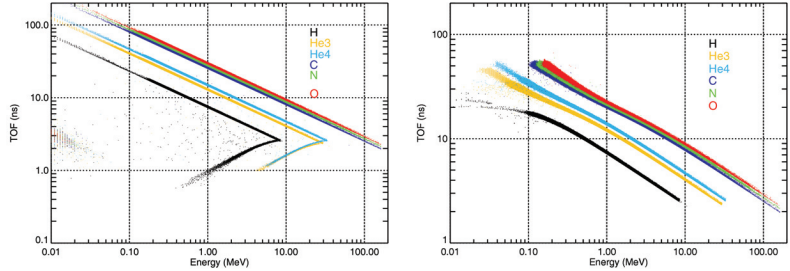

Figure 3. Simulated data in the TOF-Energy plane for different ions with a power law energy distribution.

**4.2.** *Mass resolution*

The TOF in Figure 3 (left) is calculated using a modified Eq. (1) with $L/\cos\alpha$ instead of $L$, where $\alpha$ is scattering angle to the foil normal. The energy losses in Al-dead layers are turned off. Theoretical straight tracks are reproduced for ions stopped in the SSD with a small smear in the TOF caused by scattering in the foil. For particles stopping in the SSD, the $\sigma/m$, calculated for the whole simulated energy range and for all elements is about 0.03.

Since there is no anticoincidence detector in the back of the SSD, fast light ions which are not stopped in SSD can trigger the telescope. As can be seen in Figure 3 (left), for TOF of > ~ 2.5 ns this effect does not contribute to the uncertainty in the TOF. Penetrating heavy elements storing energy above the upper threshold value in the SSD do not produce an energy signal.

In Figure 3 (right), the energy losses in the surface metallization and the dead layer were taken into account in the simulations. The simulated data in TOF-Energy plane still produce distinctive tracks although deviating from the nominal one in the low energy region. In the energy range of 0.1-1.0 MeV/n the $\sigma/m$ is ~ 0.10 for six species.

Finally, we investigate the effect of the system noise in the SIT instrument with a full width at half maximum (FWHM) of ~ 50 keV and dispersion in TOF with FWHM of ~ 1 ns. The system noise includes effects due to system electronics noise and the SSD FWHM. The uncertainties in the TOF measurement result from a combination of the TOF dispersion of secondary electrons and the noise in the MCPs and associated electronics [4]. Results are shown in Figure 4 (left). In comparison with previous TOF-Energy diagram added dispersions cause significant mass broadening, however, the helium isotopes are still distinguishable as well as oxygen from carbon ions. Mass distribution histograms in the energy range from 0.1-1.0 MeV/n for helium isotopes are shown in right panel of Figure 4. Adding effects of energy and TOF dispersions, the average mass does not change, but the mass resolution decreases to an average value of $\sigma/m \sim 0.14$ for all simulated ions. Particularly, the simulated mass resolution for $^4$He is 0.12, close to the calibration value of 0.10. The relative statistical error of the obtained masses for all elements is well below 0.01 at 95% confidence level.

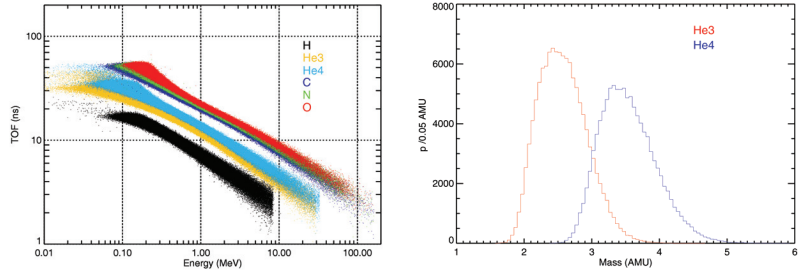

Figure 4. (Left) Same as Figure 3 with added energy and TOF dispersion. (Right) Mass histograms shown for helium $^3$He and $^4$He isotopes.

## 5. Conclusion

The Monte Carlo simulations of angular scattering and energy losses in the SIT telescope show that:
- Angular scattering in the entrance foil of the telescope is responsible for intrinsic TOF dispersion, and for the increase of the nominal instrument field-of-view which leads to high energy losses in the foil.
- Energy losses in both the SSD surface metallization and the junction dead layer contribute to the same amount to the mass resolution as the dispersion in energy and TOF measurements. Scattering in the entrance foil has a minor effect, about 30% of the previous contributors.
- Penetrating particles do not form an identifiable component of the background in the simulated instrument.

Although a further improvement in modelling of the SIT system is still needed, the present simulations can be still helpful in the experimental data analysis.

**Acknowledgments**

This work is supported by the Max-Planck-Gesellschaft zur Förderung der Wissenschaften and the Bundesministerium für Bildung und Forschung (BMBF) under grant 50 OC 0501.